\documentclass[conference]{IEEEtran}
\IEEEoverridecommandlockouts    

\usepackage{amsmath,amssymb,amsfonts}
\usepackage{algorithmic}
\usepackage{graphicx}
\usepackage{textcomp}
\usepackage{xcolor}
\usepackage[mathscr]{euscript}

\usepackage{cite}
\usepackage{tikz}

\def\BibTeX{{\rm B\kern-.05em{\sc i\kern-.025em b}\kern-.08em
    T\kern-.1667em\lower.7ex\hbox{E}\kern-.125emX}}
\newcommand\T{\rule{0pt}{2.6ex}}

\title{A Cooperative Game Theory-based Approach to Under-frequency Load Shedding Control}

\author{Mukesh Gautam, \emph{Student Member, IEEE}, Narayan Bhusal, \emph{Student Member, IEEE}, \\ and Mohammed Benidris \emph{Member, IEEE}, \\ Department of Electrical \& Biomedical Engineering, \\University of Nevada, Reno, Reno, NV 89557 \\
(emails: mukesh.gautam@nevada.unr.edu, bhusalnarayan62@nevada.unr.edu, and mbenidris@unr.edu)\vspace{-0.2ex}}

\usepackage{draftwatermark}
\SetWatermarkText{Accepted by PESGM 2021}
\SetWatermarkColor[gray]{0.1}
\SetWatermarkFontSize{0.65cm}
\SetWatermarkAngle{90}
\SetWatermarkHorCenter{20.5cm}

\begin{document}
\maketitle
%\copyrightnotice
\begin{abstract}
This paper proposes a cooperative game theory-based under-frequency load shedding (UFLS) approach for frequency stability and control in power systems. UFLS is a crucial factor for frequency stability and control especially in power grids with high penetration of renewable energy sources and restructured power systems. Conventional UFLS methods, most of which are off-line, usually shed fixed amounts of predetermined loads based on a predetermined schedule which can lead to over or under curtailment of load. This paper presents a co-operative game theory-based two-stage strategy to effectively and precisely determine locations and amounts of loads to be shed for UFLS control. In the first stage, the total amount of loads to be shed, also referred to as deficit in generation or the disturbance power, is computed using the initial rate of change of frequency (ROCOF) referred to the equivalent inertial center. In the second stage, the Shapley value, one of the solution concepts of cooperative game theory, is used to determine load shedding amounts and locations. The proposed method is implemented on the reduced 9-bus 3-machine Western Electricity Coordinating Council (WECC) system and simulated on Real-time Digital Simulators (RTDS). The results show that the proposed UFLS approach can effectively return the system to normal state after disturbances.
\end{abstract}

\section{Introduction}
%Frequency stability and control is one of the major concerns in modern power systems due to electricity market deregulation and high penetration of renewable energy sources. Frequency drop can occur when generation cannot meet system load and the system cannot maintain frequency stability with existing resources and reserve margins. Under-frequency load shedding (UFLS) schemes are usually applied if system frequency drops below a predefined frequency setting. Since load shedding is performed as the last remedial action, it should be quick, able to protect power systems from collapsing, and able to avoid unnecessary load curtailment. Also, control and protection systems of UFLS should be reliable to avoid undesirable load shedding to keep the system within its stability limits \cite{692523}. Conventionally, to retain the power balance in case of generation deficit, a fixed amount of predetermined load at predetermined locations is gradually shed based on frequency deviations.
Electric utility restructuring and increased penetration of distributed energy resources has contributed in frequency stability and control challenges in modern power systems. When existing resources and reserve margin cannot meet system load, frequency drop occurs in power systems.
If the drop in system frequency reaches a predefined threshold, under-frequency load shedding (UFLS) schemes are usually applied. Since load shedding is the ultimate way to prevent system collapse, it should be prompt, able to conserve essential and critical loads, and able to prevent the system from reaching an underfrequency condition. Also, control and protection systems of UFLS should be reliable to avoid unnecessary load shedding to keep the system within its stability limits. When designing and employing UFLS schemes, the following factors must be considered: minimum allowable frequency for secure system operation, frequency threshold for load shedding, number of load shedding locations, and load shedding amount \cite{970031}. Conventionally, under-frequency relays have been used to gradually shed a fixed amount of predetermined load based on predetermined schedule to retain balance between load and the available generation \cite{new1974load}. 

Adaptive and semi-adaptive UFLS schemes have been known to be more efficient and effective than shedding fixed amounts of predetermined loads based on predetermined schedules \cite{970031}. Out of these schemes, adaptive UFLS has the advantage of determining the minimum amount of loading for given system conditions. These schemes calculate the deficit in generation and then shed loads in adaptive steps. The computation and implementation of load shedding should be very quick to avoid frequency instability. Therefore, a reliable and quick methodology that can simultaneously determine minimum amounts of load curtailments and best locations for load shedding is pivotal for under-frequency load shedding schemes.  

Various methods have been presented in the literature for frequency control based on UFLS. An adaptive methodology to determine the minimum amount of load shedding by underfrequency relays based on the initial rate of change of frequency (ROCOF) has been proposed in \cite{May1992Anderson}. 
%An under-frequency load shedding scheme for islanded microgrids, which is not only independent of microgrid parameters but also considers power generation variations during the process, has been proposed in \cite{Nov2014Ketabi}. In \cite{Nov2016Marzband}, a stepwise load-shedding approach has been designed for islanded microgrids to regulate grid frequency while providing the amount of power shortage. 
A two-stage approach for load shedding has been presented in \cite{8323224} to address severe under frequency conditions resulting from islanding. In \cite{8416775}, a coordinated load shedding strategy based on four variable/dimensional analysis has been proposed to adaptively specify amount, location, and speed of load shedding. A two-stage UFLS scheme has been proposed in \cite{1664962} in which the amount of load shedding is estimated based on the magnitude of disturbance. 
%A first frequency derivative-based theoretically optimal UFLS scheme has been proposed in \cite{5546962}. 
A swing equation-based load-shedding scheme that maintains frequency stability of the grid with minimal load shedding has been presented in \cite{7581062}. A centralized three-stage adaptive UFLS scheme based on frequency-voltage stability has been proposed in \cite{7452538}. A sensitivity-based adaptive UFLS approach using Lagrange multipliers has been proposed in \cite{gautam2020sensitivity}. %Since optimal power flow analysis is performed in order to determine Lagrange multipliers, UFLS scheme proposed in \cite{gautam2020sensitivity} is computationally expensive and the computation time increases with the increase in system size.

Conventional UFLS schemes based on analytical modeling pose modeling and calculation complexities and their computation time increases with the increase in system size \cite{haes2019wide}. Also, they may over or under shed loads in response to frequency drop, which may lead to system instability. Cooperative game theory-based approaches have the potential to determine minimum load shedding while being computationally efficient and precise.

Game theory-based approaches (both cooperative and non-cooperative) have been successfully applied in various fields of power systems. These applications include power system reliability enhancement, loss allocation, and transmission expansion planning, to name a few. Game theory-based approaches have been used for comparing various methods of loss allocation in power systems \cite{lima2008cooperative} and for using vehicle-to-grid in frequency regulation \cite{7419270}. %and for economic dispatch \cite{Yildiran15}.
A cooperative game theory-based approach is implemented in \cite{shaloudegi2012novel} for loss reduction allocation of distributed generations using Shapley values (a solution concept in cooperative game theory). 

This paper proposes a cooperative game theory-based UFLS approach that determines optimal load shedding amounts and locations. The proposed approach distributes load shedding among various load points based on their Shapley values. The proposed approach is implemented in two-stages. In the first stage, deficit in generation is calculated based on the ROCOF in case of disturbances. In the second stage, the deficit in generation is distributed based on the equivalent Shapley values of load points. 
%In other words, the proposed method simultaneously determines locations and amounts of load-shedding steps. 
The proposed UFLS approach is applied on the reduced 9-bus 3-machine Western Electricity Coordinating Council (WECC) system and simulated on Real-time Digital Simulators (RTDS). The results show that the proposed UFLS approach can effectively return the system to normal state after contingencies.

The remainder of the paper is organized as follows. Section \ref{game} describes the cooperative game theory including the Shapley value. Section \ref{formulation} presents the formulation of the load cooperative model, which is essential for optimal load shedding. Section \ref{methodology} explains the proposed UFLS scheme for frequency control. Section \ref{cases} describes case studies on WECC-9 bus system. Section \ref{conclusion} provides concluding remarks.

\section{Cooperative Game Theory and Shapley Value}\label{game}
In game theory, a cooperative game (or coalitional game) refers to a special class of games in which players are allowed to compete with each other by forming alliances (coalitions) among themselves. Mathematically, a cooperative game is defined through specifying a certain value for each coalition. A cooperative game has the following components:
%(a) a finite set $\mathscr{N}$, and (b) a real-valued set function $V$, called characteristic function, defined on all sub-sets of $\mathscr{N}$ and satisfies $V(\phi) = 0$.
\begin{itemize}
    \item a finite set $\mathscr{N}$, and 
    \item a real-valued set function $V$, called characteristic function, defined on all sub-sets of $\mathscr{N}$ and satisfies $V(\phi) = 0$.
\end{itemize}
In the game theory terms, $\mathscr{N}$ is defined as the player set, and $V(S):2^{\mathscr{N}} \rightarrow \mathbb{R}$ is defined as the ``worth'' or ``value'' of coalition $S$, i.e., the total utility that members of $S$ can acquire if a coalition is formed among themselves and the game is played without assistance from other players.

\subsection{The Core of a Cooperative Game}
In game theory, the core refers to the set of feasible allocations that cannot be further improved through any other coalitions. Generally, outcomes of a cooperative game are expressed as n-tuples of utility: $\alpha = \{\alpha^i : i \in \mathscr{N}\}$, called payoff vectors that are measured in some common monetary unit \cite{shapley1975competitive}. 
%The set of feasible coalitionally rational payoff vectors is called the core of the game. 
The core is the set of imputations under which all sets of coalitions have values less than or equal to the sum of its members' payoffs. Thus, $\alpha$ is core if and only if,
\begin{gather}
     \alpha.e^S \geq V(S), \forall S \subset \mathscr{N} \label{stab}\\ \alpha.e^\mathscr{N} = V(\mathscr{N}) \label{eff}
\end{gather}
where $e^S$ denotes the n-vector having $e_i^S = 1$ if $i \in S$ and $e_i^S = 0$ if $i \in \mathscr{N}-S$. 
Equation \eqref{stab}, also referred to as stability or coalitional rationality, guarantees that the sum of payoffs of all players of a coalition is greater than or equal to the worth of the coalition. Equation \eqref{eff}, also referred to as efficiency, ensures that the whole amount of the grand coalition is allocated among players. 

\subsection{The Shapley Value}
The Shapley value is one solution concept of cooperative game theory. 
%The Shapley value assigns a unique payoff vector that is efficient, symmetric, and satisfies monotonicity. 
The Shapley value allocates the payoffs in such a way that is fair for cooperative solutions. 
The Shapley value of a cooperative game is given as follows \cite{curiel2013cooperative}.
\begin{equation}
    \psi_j(V) =\hspace{-1.5ex} \sum_{S \in 2^{\mathscr{N}}, j \in S}\hspace{-1.5ex} \frac{(\lvert S\rvert-1)!(n-\lvert S\rvert)!}{n!}[V(S)-V(S\backslash\{j\})]
    \label{shapley}
\end{equation}
where $n=\lvert \mathscr{N} \rvert$ is the total number of players. 

The Shapley value satisfies the following axioms:
\begin{enumerate}
    \item \textit{Efficiency:} %The efficiency axiom states that the sum of the Shapley values of all players is equal to the worth of grand coalition, so that all the gain is distributed among the players, i.e.,
    $\sum_{j \in \mathscr{N}} \psi_j(V) = V(\mathscr{N})$.
    \item \textit{Individual Rationality:} %This axiom states that the Shapley value of each player should be greater than or equal to its individual worth, i.e.,
    $\psi_j(V) \geq V(\{j\}), \forall j \in \mathscr{N}$.
    \item \textit{Symmetry:} If $j$ and $k$ are such that $V(S \cup \{j\}) = V(S \cup \{k\})$ for every coalition $S$ not containing $j$ and $k$, then $\psi_j(V) = \psi_k(V)$.
    \item \textit{Dummy Axiom:} If $j$ is such that $V(S) = V(S \cup \{j\})$ for every coalition $S$ not containing $j$, then $\psi_j(V) = 0$.
    \item \textit{Additivity:} If $V$ and $W$ are characteristic functions, then $\psi(V+W) = \psi(V) + \psi(W)$.\\
\end{enumerate}

\section{Load Cooperative Model Formulation}\label{formulation}
A load cooperative model is formulated to determine the Shapley value of each load participating in the load shedding scheme. In this paper, two types of characteristic functions are implemented for the cooperative model to capture both transient and steady-state parts of the frequency response. The first characteristic function utilizes the steady state rise in the system frequency (referred to the equivalent inertial center) resulting due to participation of loads in a particular coalition, which captures the steady state frequency response of the system. The second characteristic function utilizes the initial ROCOF for computing the worth of loads and their possible coalitions. The second characteristic function is employed in addition to the first characteristic function in order to capture the transient part of the system frequency response. Using each of these characteristic functions, two Shapley values are computed for each load using \eqref{shapley} and the equivalent Shapley value of each load is determined by taking their average.

The load cooperative model formulation for the proposed approach can be enumerated as follows.
\begin{enumerate}
    \item Collect system data including generation data, transmission line data, load data, etc., which serve as input to the load cooperative model. 
    \item Identify loads that participate in load shedding based on their priority or any other optimization techniques.
    \item Generate the list of all possible coalitions of loads. For example, if three loads ($L_1$, $L_2$, and $L_3$) are participating in load shedding, the set of all possible coalitions, denoted by $2^{\mathscr{N}}$, is as follows.
    
    $2^{\mathscr{N}} = \{\phi,\{L_1\},\{L_2\},\{L_3\},\{L_1,L_2\},\{L_1,L_3\},\\ \{L_2,L_3\},\{L_1,L_2,L_3\}\}$,\\
    where $\phi$ denotes an empty set.
    \item For each load and its possible coalitions, compute the steady state rise (due to load shedding) in system frequency with respect to the equivalent center of inertia (COI) and the initial ROCOF. The steady state rise in system frequency, $\Delta f$, and the initial ROCOF of each coalition are computed by making all the loads of a given coalition participate in ULFS. These values serve as worth of each load and their coalitions.%{\color{blue} In game theoretic terms, the disconnection the loads is equivalent to participation of those loads in UFLS.} 
    %Let $V_1$ and $V_2$, respectively denote the worth computed for each coalition (for the example in step 3) using  $\Delta f$ and the initial ROCOF. 
   % Then, the set of coalitions and their characteristic functions will be as shown in Table~\ref{tab:CharFunct}.
%\begin{table}[h!]
%\caption{Coalitions and characteristic functions for the load cooperative model}
%    \centering
%        \label{tab:CharFunct}
%    \begin{tabular}{ccc}
%    \hline
%        Possible  & Worth computed  & Worth computed using \T \\
%        coalitions&using $ \Delta f$ & initial ROCOF \T\\
%        \hline
%         $\phi$ & $0$  & $0$\T\\
%          $\{L_1\}$ & $V_1(\{L_1\})$ & $V_2(\{L_1\})$ \T\\
%          $\{L_2\}$ & $V_1(\{L_2\})$ & $V_2(\{L_2\})$ \T\\
%          $\{L_3\}$ & $V_1(\{L_3\})$ & $V_2(\{L_3\})$ \T\\
%          $\{L_1,L_2\}$ & $V_1(\{L_1,L_2\})$ & $V_2(\{L_1,L_2\})$ \T\\
%          $\{L_1,L_3\}$ & $V_1(\{L_1,L_3\})$ & $V_2(\{L_1,L_3\})$ \T\\
%          $\{L_2,L_3\}$ & $V_1(\{L_2,L_3\})$ & $V_2(\{L_2,L_3\})$ \T\\
%          $\{L_1,L_2,L_3\}$ & $V_1(\{L_1,L_2,L_3\})$ & $V_2(\{L_1,L_2,L_3\})$ \T\\
%         \hline
%    \end{tabular}\vspace{-1ex}
%\end{table}
    \item Compute two Shapley values, $\psi_i^1$ and $\psi_i^2$, of each load, $L_i$, using the characteristic functions determined in step 4 using \eqref{shapley}.
    \item Determine the equivalent Shapley value, $\psi_i^{eqv}$, of each load, $L_i$, taking the average of two Shapley values computed in step 5. The equivalent Shapley values are computed by taking the average of two Shapley values to give equal weights to both transient and steady state parts of the frequency response.
    \begin{equation}\label{eqShap}
        \psi_i^{eqv}=\frac{\psi_i^1+\psi_i^2}{2}
    \end{equation}
\end{enumerate}
The load cooperative model formulated by utilizing the procedure explained in the above steps is essential for determining locations and amounts of loads to be shed for UFLS control.

\section{The Proposed UFLS Approach}\label{methodology}
The load shedding is the fastest and ultimate way of preventing power systems from blackouts and damages initiated due to frequency drops. Load shedding is done when spinning reserves are exhausted. In this work, the proposed two-stage under-frequency load-shedding scheme is implemented in the following steps.
\begin{enumerate}
    \item Calculation of disturbance power, $P_d$, based on observed ROCOF; and
    \item Determination of locations and amounts of load shedding using the Shapley values.
\end{enumerate}

In order to calculate disturbance power $P_d$, the initial ROCOF referred to equivalent COI and the swing equation of the $i$th machine with inertia constant $H_i$ are utilized as follows \cite{May1992Anderson,5734886}.
\begin{equation} 
     (2H_i/f_n)\times (df_i/dt)=\Delta p_i
\end{equation}
\begin{equation} \label{Pd}
    P_d =\sum_{i=1}^{M}\Delta p_i=2\sum_{i=1}^{M}H_i\times(df_c/dt)/f_n
\end{equation}
where $f_n$ is the rated system frequency, $f_i$ is the frequency of the $i$\textsuperscript{th} machine, $\Delta p_i$ is the power deviation of the $i$\textsuperscript{th} machine, $M$ is the total number of machines, and $f_c$ is the frequency of the equivalent COI. 

The frequency of the equivalent COI describes the average system frequency at the time of electromechanical transients when individual machine frequencies are not the same, which can be calculated using \eqref{eq8}.
\begin{equation} \label{eq8}
     f_c =\sum_{i=1}^{M}H_i f_i \Big/ \sum_{i=1}^{M}H_i
\end{equation}

The disturbance power, $P_d$, obtained using \eqref{eq8} is distributed among candidate loads based on their equivalent Shapley values.  The locations with priority or critical loads can be removed from the list of candidate load shedding locations. Load-shedding steps are distributed among the $n$ candidate loads using distribution factor, $D_k$, defined for the $k$\textsuperscript{th} location as in \eqref{eqa}.
\begin{equation} \label{eqa}
     D_k =\psi^{eqv}_k \Big/ \sum_{k=1}^{n}\psi^{eqv}_k
\end{equation}
where $\psi^{eqv}_k$ is the equivalent Shapley value of the $k$\textsuperscript{th} load.

Now, since the sum of distribution factors of all participating loads is equal to unity, the amount of load to be shed at the $k$\textsuperscript{th} location can be calculated as follows.
\begin{equation} \label{eqb}
     p_k =D_k \times P_d
\end{equation}
The load-shedding amount, $p_k$, at the $k$\textsuperscript{th} location obtained from \eqref{eqb} may not be practical to implement. In such cases, load shedding values are rounded off in such a manner to make their sum still equal to the total disturbance power, $P_d$. 

The proposed methodology or the solution algorithm to determine the location and amount of load shedding can be summarized follows.
\begin{enumerate}
    \item Provide system data related to lines, loads, transformers, and generators.
    \item Calculate the disturbance power using initial ROCOF of equivalent COI using \eqref{Pd}.
    \item Determine the number of load-shedding locations and enumerate all possible coalitions of loads that participate in load shedding and compute their characteristic functions. 
    \item For each load-shedding location, compute the equivalent Shapley value of each load based on \eqref{eqShap} and the respective amount of load to be shed using \eqref{eqb}. For example, if load-shedding is to be distributed among three loads, the equivalent Shapley values and the amounts of loads to be shed are determined for each of those three loads.
\end{enumerate}

\begin{figure}
    \centering
    \vspace{-3.5ex}
    \includegraphics{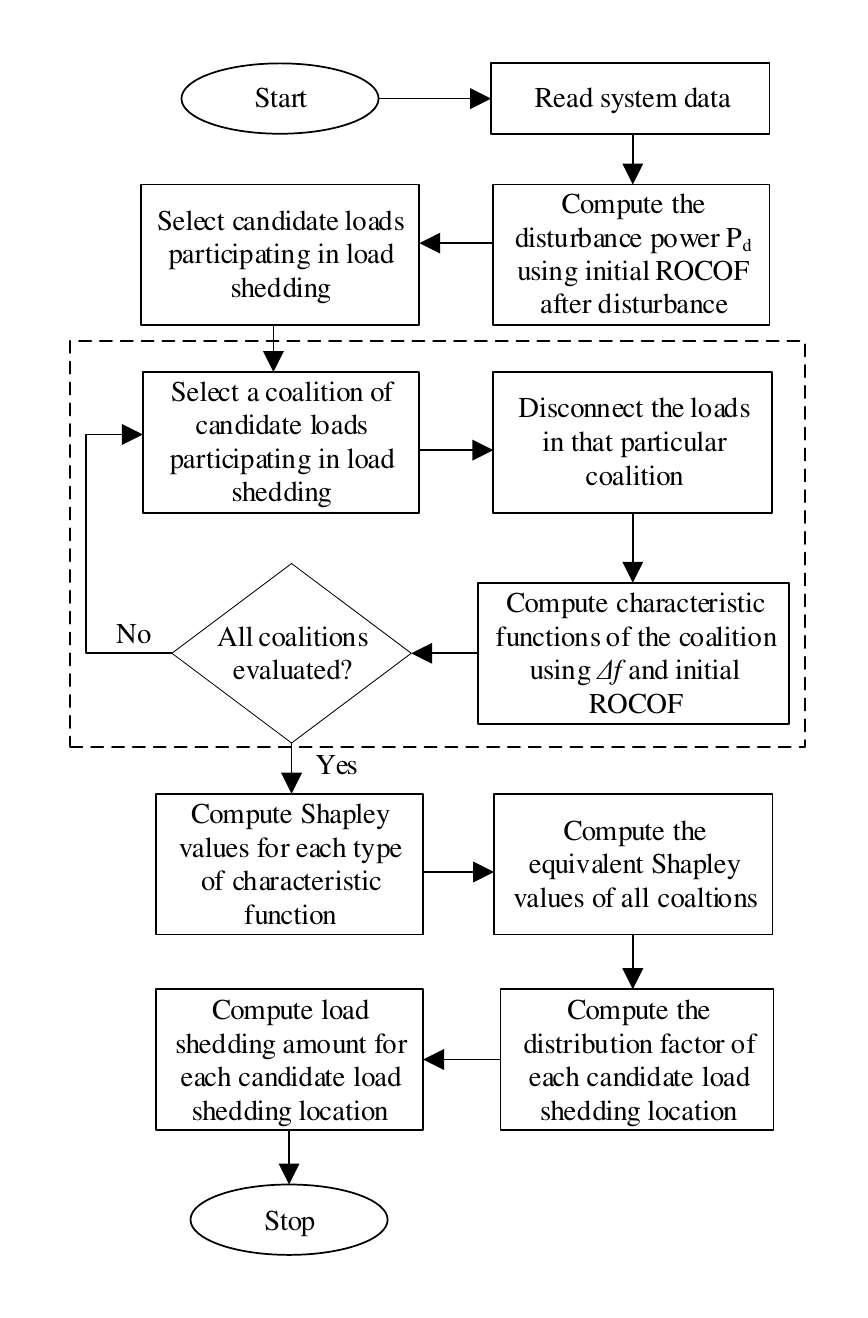}\vspace{-4ex}
    \caption{Flow chart of proposed UFLS scheme}
    \label{fig1}
\end{figure}

The flowchart of the proposed solution algorithm is shown in Fig. \ref{fig1}.  The steps for calculating characteristic functions of each coalition (shown inside dashed rectangle in the flowchart) can be done offline. Since the calculation of characteristic functions can be done off-line, the proposed UFLS approach can be applied for primary frequency control as it only requires the calculation of initial ROCOF based on the magnitude of the disturbance.

\section{Case Study and Discussions}\label{cases}
The proposed method is implemented on the reduced WECC 9-bus system. This system consists of three generators and three load points with total loading of 315 MW and 115 MVar. System data and configuration are provided in \cite{AndersonFaud77} and inertia constants are given in Table \ref{WECC}. This system has been extensively used in several power system stability studies.
\begin{table}[]
\caption{Machine inertia data for WECC-9 bus system\vspace{-1.5ex}}
\centering
\begin{tabular}{|c|c|}
\hline
\textbf{Machine number} & \textbf{Inertia constant (H)} \T\\ \hline
1                       & 23.64                         \T\\ \hline
2                       & 6.4                           \T\\ \hline
3                       & 3.01                          \T\\ \hline
\end{tabular}
\label{WECC}
\end{table}

In order to compute the worth (or characteristic function) for each load and their possible coalitions in the tested system, the steady state rise  in the system frequency referred to the equivalent COI, $\Delta f$, and the initial ROCOF are computed through dynamic simulations using RTDS.  When the load of bus $5$ is allowed to participate in UFLS, the steady-state rise in system frequency and initial ROCOF are, respectively, $1.2757$ Hz and $1.1189$ Hz/s. Similarly, if both loads at buses $5$ and $6$ are allowed to participate in UFLS, the steady-state rise in system frequency and the initial ROCOF are, respectively, $2.1869$ Hz and $1.9169$ Hz/s. Table~\ref{tab:coalitions} shows the characteristic functions of possible coalitions in terms of $\Delta f$ and initial ROCOF for all possible coalitions for this system.
\begin{table}
\caption{Characteristic Function of Possible Coalitions in Terms of $\Delta f$ and Initial $ROCOF$\vspace{-1.5ex}}
\centering
\begin{tabular}{|c|c|c|}
\hline
Load bus coalitions & $\Delta f$ (Hz) & Initial $ROCOF$ (Hz/s) \T\\ \hline
5                   & 1.2757                       & 1.1189               \T\\ \hline
6                   & 0.9196                       & 0.7990               \T\\ \hline
8                   & 0.9887                       & 0.8890               \T\\ \hline
5 \& 6              & 2.1869                       & 1.9169               \T\\ \hline
5 \& 8              & 2.2727                       & 2.0092               \T\\ \hline
6 \& 8              & 1.9144                       & 1.6866               \T\\ \hline
5, 6 \& 8           & 3.1883                       & 2.8071               \T\\ \hline
\end{tabular}
\label{tab:coalitions}
\end{table}

For the reduced WECC 9-bus system, the outage of machine-3 is simulated. As there are three load points in the system, system operators have to determine a certain number (or steps) of load shedding locations and amounts. Therefore, in this paper, cases of two and three load-shedding locations are considered. For the disturbance power of 85 MW, if two locations are chosen for load shedding, the results show that $49$ MW is to be shed from bus $5$ and $36$ MW from bus $8$. The load-shedding amount along with the values of equivalent Shapley values for different loads are given in Table \ref{tab:twoStep}. 
\begin{table}[h!]
\caption{Locations and Sizes of Two Step Load Shedding\vspace{-1.5ex}}
    \centering
    \begin{tabular}{c|c|c}
    \hline
        Load-shedding  & Equivalent Shapley & Load-Shedding \T \\
        Locations & values & Steps \\
        \hline
         Bus $5$ & $1.2222$  & $49$ MW \T\\
         Bus $8$ & $0.9187$ & $36$ MW \T\\
         \hline
    \end{tabular}
    \label{tab:twoStep}
\end{table}

% \begin{figure}
% %    \centering
% \hspace{-2.5ex}
%     \includegraphics[scale=0.28]{WECC.pdf}
%     \caption{Three-machine test power system}
%     \label{ThreeMachine}
% \end{figure}

If three load shedding locations are chosen for the same disturbance power of $85$ MW, the following amounts of loads are shed from buses $5$, $6$, and $8$ respectively: $34$ MW, $24$ MW, and $27$ MW. The load-shedding amounts and the values of equivalent Shapley values for a three-step load-shedding are as shown in Table \ref{tab:Threestep}.
\begin{table}[h!]
\caption{Locations and Sizes of Three Step Load-Shedding\vspace{-1.5ex}}
    \centering
    \begin{tabular}{c|c|c}
    \hline
        Load-shedding  & Equivalent Shapley & Load-Shedding \T \\
        Locations&values &Steps \T\\
        \hline
         Bus $5$ & $1.1973$  & $34$ MW \T\\
         Bus $6$ & $0.8581$ & $24$ MW \T\\
         Bus $8$ & $0.9424$  & $27$ MW \T\\
         \hline
    \end{tabular}\vspace{-1ex}
    \label{tab:Threestep}
\end{table}

\begin{figure}
    \centering
    \includegraphics[scale=0.6]{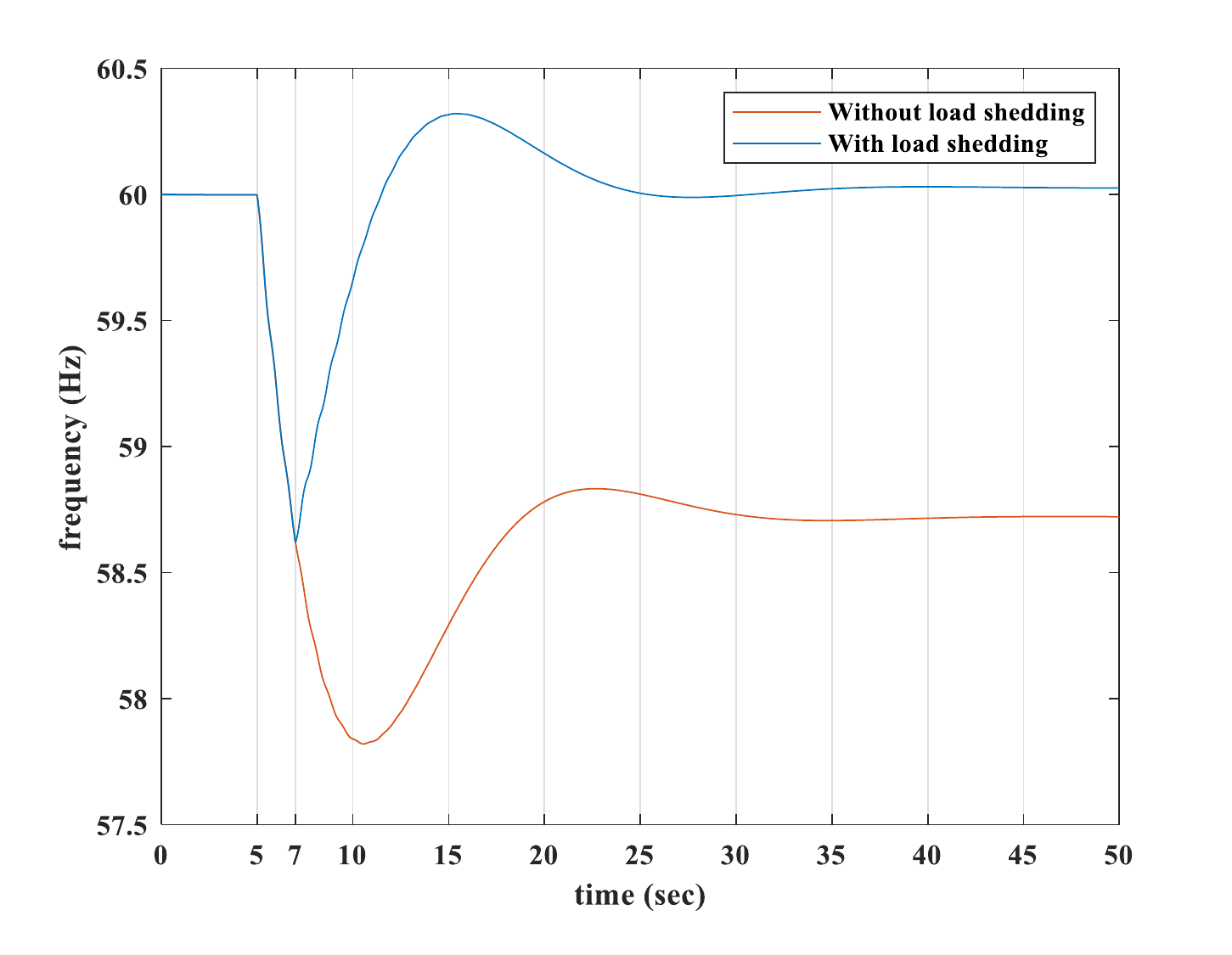}\vspace{-2ex}
    \caption{Plot of system frequency}
    \vspace{-1ex}
    \label{freqplot}
\end{figure}

In order to study system dynamics with and without load-shedding, the WECC 9-bus system is simulated on RTDS considering the outage of machine-2 at second 5 and load-shedding after two seconds of the outage. The three-step load-shedding amounts as shown in Table~\ref{tab:Threestep} are used for dynamic simulations on the RTDS. The plot of system frequency (without and with load-shedding) of the reduced WECC 9-bus system, which are simulated on the RTDS, is shown in Fig. \ref{freqplot}. From the frequency response we can see that without load shedding, the system frequency drops to nearly $57.7$ Hz at second 10.5, where steady-state frequency being nearly $58.6$ Hz. However, when load shedding is performed at second 7 (i.e., two seconds after the outage), the system frequency recovers to the normal frequency level.
The frequency plots in Fig. \ref{freqplot} show that the load-shedding is effective in returning the system to normal frequency.

The proposed UFLS approach is compared with the sensitivity-based approach presented in \cite{gautam2020sensitivity}. When the study is performed on a PC with 64-bit Intel i5 core, 3.15 GHz processor, 8 GB RAM, and Windows OS, the sensitivity-based approach presented in \cite{gautam2020sensitivity} takes $0.23$ seconds for the reduced WECC 9-bus system and $180.36$ seconds for the New England 39-bus system to distribute the load shedding amounts. On the other hand, the proposed UFLS approach can distribute the load shedding amounts within a fraction of milliseconds irrespective of the system size.

\section{Conclusion}\label{conclusion}
In this paper, a cooperative game theory-based two-stage UFLS scheme has been proposed. In the first stage, the computation of disturbance power was performed using initial ROCOF. In the second stage, equivalent Shapley values were determined for the loads that are participating in load shedding. In order to calculate the equivalent Shapley values of the loads, the characteristic functions of loads and their possible coalitions were calculated using steady-state rise in system frequency and inital ROCOF evaluated through dynamic simulation of the system. Using the equivalent Shapley values of loads, the disturbance power was distributed among different loads. Through the dynamic simulations of WECC-9 bus system using RTDS, it was shown that the proposed methodology can be effectively implemented for frequency control of power systems.  

\section*{Acknowledgement}
This work was supported in Part by the U.S. Department of Energy (DOE) under Grant DE-EE0009022. This work was prepared as an account of work sponsored by an agency of the United States Government. Neither the United States Government nor any agency thereof, nor any of their employees, makes any warranty, express or implied, or assumes any legal liability or responsibility for the accuracy, completeness, or usefulness of any information, apparatus, product, or process disclosed, or represents that its use would not infringe privately owned rights.

\bibliographystyle{IEEEtran}
\bibliography{References.bib}

%\appendices
%\section{Data Used in Simulation}
%\label{APX}

\end{document}